\documentclass[12pt]{article}
\usepackage{graphicx, lineno}
\begin{document}
	
%
%
%

\begin{center}
    \sf {\Large {\bfseries On the calculation of the radiobiological effect of radiolytic oxygen depletion in FLASH radiotherapy}} \\
    \vspace*{10mm}
    Juan Pardo-Montero$^{1,2,*}$, Isabel González-Crespo$^{1,+}$ \\
    $^1$Group of Medical Physics and Biomathematics, Instituto de Investigación Sanitaria de Santiago (IDIS), Santiago de Compostela, Spain \\
     $^2$Department of Medical Physics, Complexo Hospitalario Universitario de Santiago, Santiago de Compostela, Spain
    \vspace{5mm}\\
    Version typeset \today\\
\end{center}

\pagenumbering{roman}
\setcounter{page}{1}
\pagestyle{plain}

* E-mail:  juan.pardo.montero@sergas.es \\

+ Now at Institute of Analysis and Scientific Computing, Technische Universität Wien, Vienna, Austria.

	
\begin{abstract}

\noindent\textbf{Objective}: Radiolytic oxygen depletion (ROD) may play a role in the sparing of cells irradiated with ultra-high dose rates. Different methods have been used to quantify the effect of ROD during FLASH irradiation on cell survival, typically involving some kind of averaging of the oxygen effect  and the LQ model. In this work, we compare the results obtained with several of these methods and introduce a novel method based on the non-linear differential form of the LQ model.

\noindent\textbf{Approach}: We present a novel method to account for a varying oxygen concentration on the dose-response based on the non-linear differential form of the LQ model, and we compare the results obtained with this method with those obtained with other methods that linearize the averaging of the oxygen effect during irradiation.

\noindent\textbf{Main results}: We found differences in the surviving fractions obtained with the method introduced in this work and other methods that introduce different linearizations (averaging) of the non-linear dependence on the oxygen concentration, especially for oxygenations and doses that lead to important changes in the OERs during the delivery of the dose (initial oxygenations $\approx$5--10 mmHg and doses $>30$~Gy). On the other hand, we showed that the method presented by Zhu \emph{et al.} is equivalent to a first-order Euler numerical method of the differential LQ model.

\noindent\textbf{Significance}: The method introduced in this work and the method of Zhu \emph{et al.} may allow a more precise quantification of the effect of ROD on dose-response, both for tumors and normal tissues. While all the reviewed methods show an oxygen-dependent sparing effect of FLASH radiotherapy driven by ROD and qualitatively similar results, the method introduced in this work and that of Zhu \emph{et al.} may be more suitable to quantitatively analyze new preclinical (and future clinical) data coming from experimental studies. 
\end{abstract}

%
\vspace{2pc}

\section{Introduction} \label{section_intro}

The delivery of radiotherapy at ultra-high dose rates has generated significant interest in recent years, as \textit{in vivo} experiments have shown its potential to spare non-tumor tissue while seemingly maintaining the effectiveness of conventional radiotherapy, known as the FLASH effect~\cite{favaudon2014, montay2017, vozenin2019}.

It is not entirely clear yet what the mechanism behind the FLASH effect is, and it may arise from a combination of different biological and physicochemical effects. Among the investigated factors are radiation-induced immune effects~\cite{jin2020}, physicochemical differences in the production yield and recombination of free radicals~\cite{spitz2019, abolfath2022, shukla23}, and the protective effect of radiolytic oxygen depletion (ROD) caused by FLASH radiotherapy~\cite{favaudon2022, pratx2019b, petersson2020}. The latter has probably been the most investigated as the cause behind the FLASH effect: ROD has been experimentally observed during ultra-high dose rate irradiation both \emph{in vitro} and \emph{in vivo}~\cite{cao2021, ha2022, van2022, karle2024}, and it is well known from classical radiobiology that lower oxygen levels increase the radioresistance of cells~\cite{wouters1997}. While the ROD effect observed in experiments may not be large enough to account for the important sparing of healthy tissues/organs achieved with FLASH radiotherapy \cite{limoli2023, vozenin2022}, ROD continues to be actively investigated as a cause of the FLASH effect \cite{tavakkoli2025}, and also as a potential cause of tumor sparing \cite{liew2023, gonzalez2024}.

The quantification of the effect of a variable oxygen concentration during dose delivery on the surviving fraction of cells requires some method to average the radiobiological effect of the oxygen dynamics. While some authors rely on modeling the dynamics of damage \cite{liew2023}, most methods presented in the literature rely on the averaging of the oxygen-dependent linear-quadratic (LQ) response of cells to radiation, including the averaging of surviving fractions, effective $\alpha$ and $\beta$ parameters, or oxygen enhancement ratios (OER). In this work, we investigate the surviving fractions obtained with different methods versus dose and oxygenation, namely the methods presented in \cite{taylor2022, gonzalez2024} (averaging of the surviving fraction), \cite{song2023} (averaging of radiosensitivity parameters), and \cite{petersson2020} (averaging the oxygen enhancement ratios). We also investigate the iterative method presented in \cite{zhu2024}, which does not explicitly include an averaging of the oxygen effect. Finally, we present a novel method to account for a varying oxygen concentration on the dose-response based on the non-linear differential form of the LQ model.

\section{Materials and Methods} \label{section_materials}

\subsection{Radiolytic oxygen depletion in FLASH radiotherapy} \label{section_materials_pO2}

For spatially heterogeneous oxygen distributions like those present \emph{in vivo} we followed the work of Taylor~\textit{et~al.} \cite{taylor2022}, also implemented in González-Crespo \textit{et al.} \cite{gonzalez2024}, and modeled the dynamics of oxygen, $p(t,\mathbf{x})$, with a spatiotemporal reaction-diffusion equation:

\begin{equation}
     \frac{\partial p(\mathbf{x},t)}{\partial t} = D_\mathrm{O_2} \Delta p(\mathbf{x},t) - g_\mathrm{max}\frac{p(\mathbf{x},t)}{k+p(\mathbf{x},t)}-\frac{D}{T}G_0\frac{p(\mathbf{x},t)}{k_\mathrm{ROD}+p(\mathbf{x},t)}, \label{eq_oxyFlash}
\end{equation}

In this equation $t$ and $\mathbf{x}$ are the time and spatial coordinates. The first term models oxygen diffusion, with $D_\mathrm{O_2}$ being the diffusion coefficient of oxygen and $\Delta$ the Laplacian operator. The next two terms have the form of Michaelis-Menten kinetics and model oxygen depletion due to both metabolism and radiolysis, with $g_\mathrm{max}$ the maximum metabolic consumption rate, $k$ the oxygen pressure for half-maximum consumption rate, $D$ the radiation dose, $T$ the total time of irradiation, $G_0$ the radiolytic consumption rate, and $k_{\rm ROD}$ the oxygen pressure for half-maximum oxygen depletion rate. In addition, the capillaries, heterogeneously distributed, act as oxygen sources with $p$=40 mmHg, as described in detail in \cite{gonzalez2024} and references therein.

We also considered a simpler situation, where oxygenation is supposed to be homogeneous and there is no diffusion, emulating an \emph{in vitro} scenario. In this situation, we have simplified the dynamics of oxygen to 
\begin{equation}
     \frac{{\rm d} p(t)}{{\rm d} t} = -\frac{D}{T}G_0\frac{p(\mathbf{x},t)}{k_\mathrm{ROD}+p(\mathbf{x},t)}, \label{eq_oxyFlash_2}
\end{equation}
ignoring the contributions of consumption and recovery because those terms are much smaller than radiolytic depletion during the delivery of the dose.

\subsection{Calculation of the effect of variable oxygenation during dose delivery in FLASH-RT}

We relied on the LQ model \cite{fowler1989} to describe the surviving fraction, $\mathit{SF}$, of cells with oxygenation $p$ irradiated to a dose $D$:

\begin{equation}
    \mathit{SF}(p,D) = \exp (-\alpha(p) D-\beta(p) D^2).
    \label{eq_LQ}
\end{equation}
The $\alpha$ and $\beta$ parameters depend on the oxygen partial pressure as \cite{wouters1997}:
\begin{eqnarray}
	&\displaystyle \mathit{\alpha(p)} = \frac{\alpha_{\mathrm{ox}}}{{\mathit{M}_\alpha}}\frac{{\mathit{M}_\alpha} p  + k_\mathrm{m}}{p  + k_\mathrm{m}}, \label{eq_OERa}\\
	&\displaystyle \mathit{\beta(p)} = \frac{\beta_{\mathrm{ox}}}{{\mathit{M}_\beta}^2}\frac{({\mathit{M}_\beta} p  + k_\mathrm{m})^2}{(p  + k_\mathrm{m})^2},
	\label{eq_OERb}
\end{eqnarray}
where $\alpha_\mathrm{ox}$ and $\beta_\mathrm{ox}$ are the $\alpha$ and $\beta$ parameters under fully aerobic conditions, ${\mathit{M}_\alpha}$ and  ${\mathit{M}_\beta}^2$ are the maximum oxygen enhancement ratios for $\alpha$ and $\beta$, respectively, and $k_\mathrm{m}$ is the oxygen partial pressure at which the OERs equal the half-maximum value.

Different methods have been used to quantify the effect of variable oxygenation during FLASH irradiation and obtain the surviving fraction after a FLASH-RT irradiation, $\mathit{SF}_\mathrm{F}$. In this work, we investigate four methods from the literature, as well as a novel method based on the differential formulation of the LQ model. The dose rate, $R$, of the FLASH-RT delivery is considered constant during the irradiation time.


\begin{itemize}

\item{{Method 1: averaging the surviving fraction.}}


In Taylor~\textit{et~al.} and González-Crespo \textit{et al.}~\cite{taylor2022, gonzalez2024} the effect of a FLASH-RT treatment was calculated by averaging the surviving fraction during the irradiation time $T$ as:

\begin{equation}
\mathit{SF}_\mathrm{F} = \frac{1}{T} \int_{0}^{T}\mathit{SF}(p(t),D) \mathrm{d}t,
	\label{eq_SF_model1}
\end{equation}
where $p(t)$ is the time-dependent oxygenation, which is obtained by solving Eqs.~(\ref{eq_oxyFlash}) or (\ref{eq_oxyFlash_2}), and the surviving fraction $\mathit{SF}(p(t),D)$ is given by Eqs.~(\ref{eq_LQ}), (\ref{eq_OERa}), and (\ref{eq_OERb}).

\item{Method 2: averaging $\alpha$ and $\beta$.}

A different approach was used in Song~\textit{et~al.}~\cite{song2023}, where the average $\alpha$ and $\beta$ during irradiation were calculated as:

\begin{eqnarray}
\bar{\alpha} &=& \frac{1}{T} \int_{0}^{T} \alpha (p(t)) dt = \frac{1}{T} \int_{0}^{T} \frac{\alpha_{\mathrm{ox}}}{{\mathit{M}_\alpha}}\frac{{\mathit{M}_\alpha} p(t)  + k_\mathrm{m}}{p(t)  + k_\mathrm{m}} dt, \label{eq_a_av}\\
\bar{\beta} &=& \frac{1}{T} \int_{0}^{T} \beta (p(t)) dt = \frac{1}{T} \int_{0}^{T} \frac{\beta_{\mathrm{ox}}}{{\mathit{M}_\beta}^2}\frac{({\mathit{M}_\beta} p(t)  + k_\mathrm{m})^2}{(p(t)  + k_\mathrm{m})^2} dt ,	\label{eq_b_av}
\end{eqnarray}
and from these values, the surviving fraction is calculated as $\mathit{SF}_\mathrm{F}$:

\begin{equation}
\mathit{SF}_\mathrm{F} = \exp (-\bar{\alpha} D- \bar{\beta} D^2).
	\label{eq_SF_model2}
\end{equation}

\item{Method 3: averaging the oxygen enhacement ratios.}

Petersson~\textit{et~al.}~\cite{petersson2020} averaged the oxygen enhancement ratio ($\mathit{OER}$) during the irradiation. The averaged $\mathit{OER}$ is calculated as:

\begin{equation}
\overline{\mathit{OER}} = \frac{1}{T} \int_{0}^{T} \mathit{OER} (p(t)) dt = \frac{1}{T} \int_{0}^{T} \left ( \frac{M p(t) + k_\mathrm{m}}{p(t)+k_\mathrm{m}} \right ) \mathrm{d}t, \label{eq_OER_av}
\end{equation}
where $M$ refers to $M_\alpha$ or $M_\beta$ in Eqs.~(\ref{eq_OERa}) and (\ref{eq_OERb}). With the values of $\overline{\mathit{OER}}$ for $\alpha$ and $\beta$, the surviving fraction $\mathit{SF}_\mathrm{F}$ can be calculated as:

\begin{equation}
\mathit{SF}_\mathrm{F} = \exp \left( -\alpha_{\rm ox} \frac{\overline{\mathit{OER_\alpha}}}{M_\alpha} D- \beta_{\rm ox} \frac{\overline{\mathit{OER_\beta}}^2} {M_\beta ^2} D^2 \right ).
	\label{eq_SF_model3}
\end{equation}

Notice that if the averaging is instead applied to the quadratic expression for the $\beta$-term,

\begin{equation}
\overline{\mathit{OER}_\beta ^2} = \frac{1}{T} \int_{0}^{T} \left ( \frac{M_\beta p(t) + k_\mathrm{m}}{p(t)+k_\mathrm{m}} \right )^2 \mathrm{d}t, \label{eq_OER2_av}
\end{equation}
the expression for the surviving fraction becomes:
\begin{equation}
\mathit{SF}_\mathrm{F} = \exp \left( -\alpha_{\rm ox} \frac{\overline{\mathit{OER_\alpha}}}{M_\alpha} D- \beta_{\rm ox} \frac{\overline{\mathit{OER_\beta^2}}} {M_\beta ^2} D^2 \right ),
	\label{eq_SF_model3p}
\end{equation}
which is identical to Model 2.

\item{ Method 4: Zhu's iterative method.}

Zhu~\textit{et~al.}~\cite{zhu2024} presented an iterative method to calculate the effect of ROD on the surviving fraction. The radiation dose is discretized in steps $D_n$=$n \times \Delta D$ ($0 \le n \le n_M$). At each step $n$ the radiosensitivity parameters can be calculated according to the oxygenation levels at the time the radiation dose $D_n$ has been delivered according to Eqs.~(\ref{eq_OERa}) and (\ref{eq_OERb}), resulting in values that we denote as $\alpha_n$ and $\beta_n$. Using our notation, the surviving fraction can be calculated as:

\begin{equation}
\left\{
\begin{array}{ll}
                  \mathit{SF}_{\mathrm{F},0}=1 & \textnormal{for $n$=0},\\
                  \mathit{SF}_{\mathrm{F},n}=\mathit{SF}_{\mathrm{F},n-1} \times d_{SF,n} & \textnormal{for $0 < n \le n_M$},\\
                \end{array}
              \right.
              \label{zhu}
\end{equation} 
where $D_{n_M}=D$, and  
\begin{equation}
d_{SF,n}=\frac{\exp(-\alpha_n D_n - \beta_n D_n^2)}{\exp(-\alpha_n D_{n-1} - \beta_n D_{n-1}^2)}.
\end{equation} 

While Zhu~\textit{et~al.}~\cite{zhu2024} originally used a discretization step $\Delta D$= 1~Gy, the method can be employed with different discretizations.

\item{Method 5: differential LQ method.}

Methods 1--3 include a linearization through the averaging integral of a non-linear effect. In this work, we have used the differential form of the LQ model~\cite{scheidegger2011}, which allows a more natural evaluation of the radiobiological effects of a dynamic oxygenation. Assuming a constant dose rate, $R$, during the delivery, the evolution of the surviving fraction with time can be written as:

\begin{equation}
\frac{ {\rm d} \mathit{SF} (t) } {{\rm d}t}  = - \left [ \alpha(p(t)) R + 2 \beta(p(t)) R d(t)\right ] \mathit{SF}(t),
\label{LQ_diff}
\end{equation}
where $d(t)$ is the dose delivered up to time $t$ ($d(t)=Rt$).

This equation can be numerically solved for general oxygenation curves $p(t)$ with the initial condition $\mathit{SF}(t=0)=1$. The surviving fraction after the irradiation is given by $\mathit{SF}_\mathrm{F}\equiv \mathit{SF}(t=T)$ (with $T$ the time taken to deliver the dose $D$; $D=RT$).\footnote{In a more general case where the dose rate is not constant during delivery, we can write this equation as $ {\rm d} \mathit{SF} (t) / {\rm d}t  = - \left [ \alpha(p(t)) R(t) + 2 \beta(p(t)) R(t) (\int_0^t R(\tau) \mathrm{d}\tau)\right ] \mathit{SF}(t)$, and solve it with the same method. However, we will restrict our analysis to a constant dose rate during delivery as this is the case usually considered in modeling studies.}

\end{itemize}

In addition to computing FLASH surviving fractions with the different methods, we have also computed the surviving fractions for conventional radiotherapy, $\mathit{SF}_\mathrm{C}$, for the same doses and oxygenations. These were computed by assuming that there is no oxygen depletion during dose delivery, therefore the oxygen partial pressure is constant and $\mathit{SF}_\mathrm{C}$ is given by Eqs.~(\ref{eq_LQ})--(\ref{eq_OERb}).

\subsection{Parameter values}

We investigated differences in computed $\mathit{SF}_\mathrm{F}$ values for different doses, $D$, and oxygenations, $p$. The rest of the parameters were fixed to values usually employed in the radiobiological literature, and are reported in Table \ref{table_0}. We included both homogeneous and heterogeneous oxygenation distributions. Regarding the latter, two oxygen distributions obtained from two different vascular fractions and capillary distributions were taken from González-Crespo \emph{et al} \cite{gonzalez2024}, corresponding to poorly and moderately well oxygenated tissues.

\begin{table}[htb]
    \caption{\label{table_0} List of parameter values used in this work.}
    \begin{center}
        \begin{tabular}{@{}ll}
            \hline
    	Parameter              & Value            \\
            \hline
           $D_\mathrm{O_2}$        & 2$\times$10$^{-9}$ m$^2$ s$^{-1}$ \\
           $g_\mathrm{max}$        & 15 mmHg s$^{-1}$ \\
           $k$                     & 2.5 mmHg         \\
           $k_{\rm ROD}$           & 1 mmHg        	\\
           $G_0$                   & 0.25 mmHg Gy$^{-1}$ \\
           $\alpha_{\rm ox}$       & 0.3 Gy$^{-1}$     \\
           $\beta_{\rm ox}$       & 0.03 Gy$^{-2}$         \\
           $M_\alpha$ & 2.5              			\\
           $M_\beta$  & 2.5              \\
           $k_\mathrm{m}$          & 3.28 mmHg        \\
           $R$          & 100 Gy s$^{-1}$        \\
           \hline
        \end{tabular}
    \end{center}
\end{table}


\subsection{Implementation and solvers}

The models and methods were implemented in MATLAB R2022b (Natick, USA). The averaging integrals in equations Eqs.~(\ref{eq_SF_model1}), (\ref{eq_a_av}), (\ref{eq_b_av}), (\ref{eq_OER_av}) and (\ref{eq_OER2_av}) were solved with a standard numerical quadrature method. The differential equation (\ref{LQ_diff}) was solved with a fourth-order Runge-Kutta method (RK4)~\cite{butcher2016} (we observed that lower order methods, like Euler's, were not accurate enough). The implementation of the RK4 method is as follows:

\begin{equation}
\left\{
\begin{array}{r@{\quad}l}
  & t_0=0\\
 &\mathit{SF}_0=1  \\
  &  t_{n+1}=t_n+h \\
  &\mathit{SF}_{n+1} = \mathit{SF}_n + \frac{h}{6}(k_1+2k_2+2k_3+k_4)
\end{array}
\right.
\end{equation}
where $n=0, \cdots, n_M$, $h=T/n_M$, and,

\begin{equation}
\left\{
\begin{array}{r@{\quad}l}
& k_1=f(t_n, \mathit{SF}_{n}, \alpha_n, \beta_n)\\
& k_2=f(t_n+ h/2, \mathit{SF}_{n} + k_1 h/2, (\alpha_n+\alpha_{n+1})/2, (\beta_n+\beta_{n+1})/2) \\
& k_3=f(t_n+ h/2, \mathit{SF}_{n} + k_2 h/2, (\alpha_n+\alpha_{n+1})/2, (\beta_n+\beta_{n+1})/2) \\
& k_4=f(t_n+ h, \mathit{SF}_{n} + k_3 h, \alpha_{n+1}, \beta_{n+1}) \\
\end{array}
\right.
\end{equation}
with,

\begin{equation}
f(t, \mathit{SF}, \alpha, \beta)= - (\alpha R + 2 \beta R^2 t ) \mathit{SF}.
\end{equation}

\section{Results and Discussion} \label{section_results}

\subsection{The effect of ROD for homogeneous oxygenations}
In this section, we use $\mathit{SF}_{\mathrm{F}_1}$, $\mathit{SF}_{\mathrm{F}_2}$, $\mathit{SF}_{\mathrm{F}_{3}}$, $\mathit{SF}_{\mathrm{F}_4}$, and $\mathit{SF}_{\mathrm{F}_5}$ to refer to the surviving fractions computed by using Eqs.~(\ref{eq_SF_model1}), (\ref{eq_SF_model2}), (\ref{eq_SF_model3}), (\ref{zhu}) and (\ref{LQ_diff}), respectively. We quantified differences among methods as:

\begin{equation}
\mathit{diff}_{5, i}=| \log_{10} \mathit{SF}_{\mathrm{F}_5}(p, D) - \log_{10} \mathit{SF}_{\mathrm{F}_i}(p, D)|
\label{comparison}
\end{equation}
using the differential method (method 5) as reference.

\begin{figure}[]
\centering
\includegraphics[width=\columnwidth]{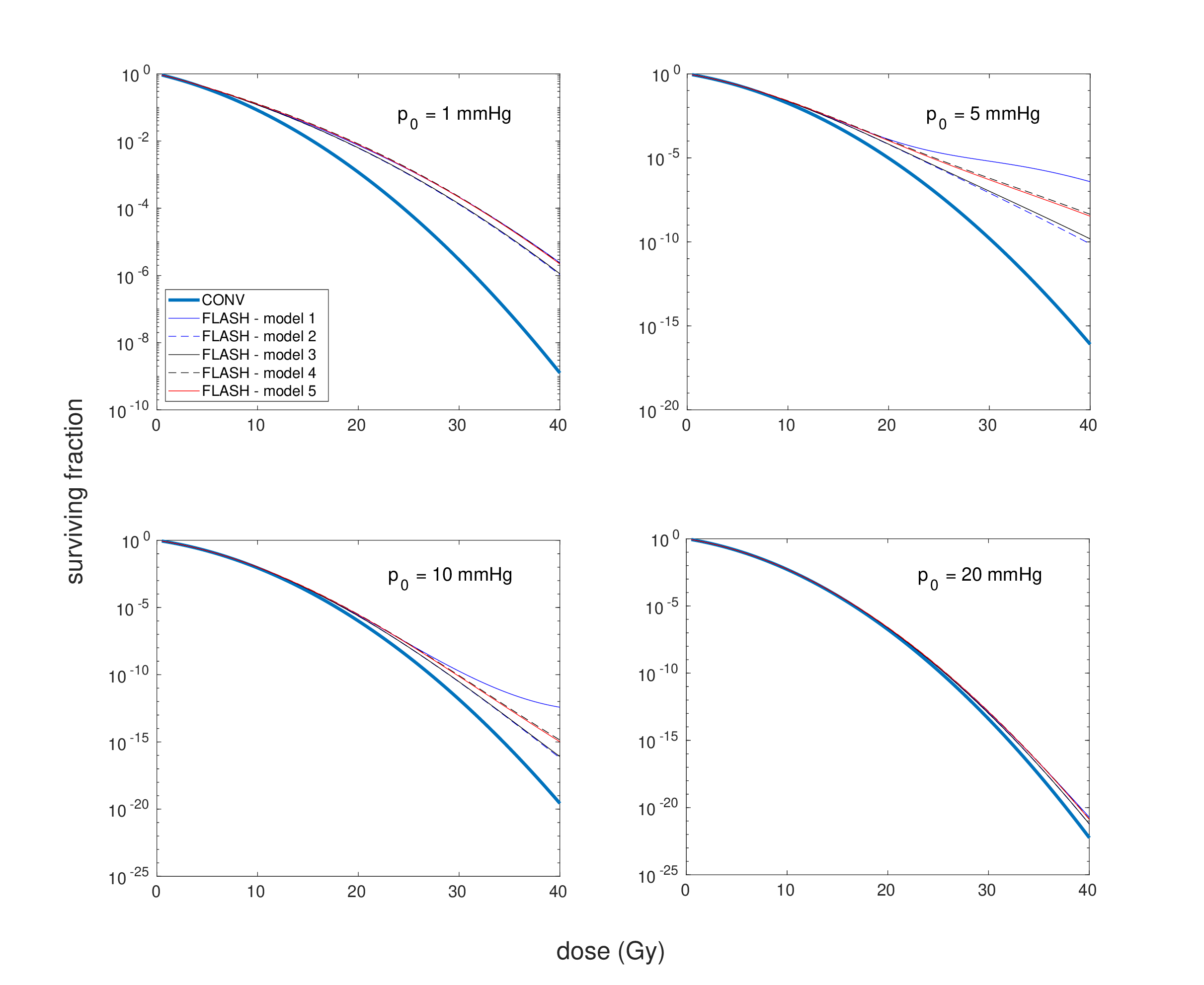}
\caption{\small Surviving fractions versus dose at different initial oxygenation levels (1, 5, 10 and 20 mmHg) for CONV-RT (solid line) and FLASH-RT computed with different methods to account for the effect of oxygen depetion: Method 1, averaging of the surviving fraction; Method 2, averaging of $\alpha$ and $\beta$; Method 3, averaging of OERs; Method 4, Zhu's iterative method; and Method 5, the differential LQ method.}
\label{fig1}
\end{figure}

In Figure \ref{fig1} we report surviving fractions computed with the different methods employed in this work for homogeneously oxygenated cells with different levels of oxygenation (1, 5, 10 and 20~mmHg) irradiated with FLASH doses ranging from 0.5~Gy to 40~Gy. All methods yield surviving fractions that are higher than those of conventional therapy, as expected from the oxygen depletion effect. All methods yield very similar results for very low oxygenations, where there is little oxygen to deplete, and for well oxygenated cells, where the effect of oxygen depletion does not change the radiosensitivity of the cells significantly. However, for moderately oxygenated cells (5--10 mmHg), there are important differences among the different methods, especially for Method 1 (averaging of the surviving fraction) which leads to higher surviving fractions.

\begin{figure}[]
	\centering
	\includegraphics[width=\columnwidth]{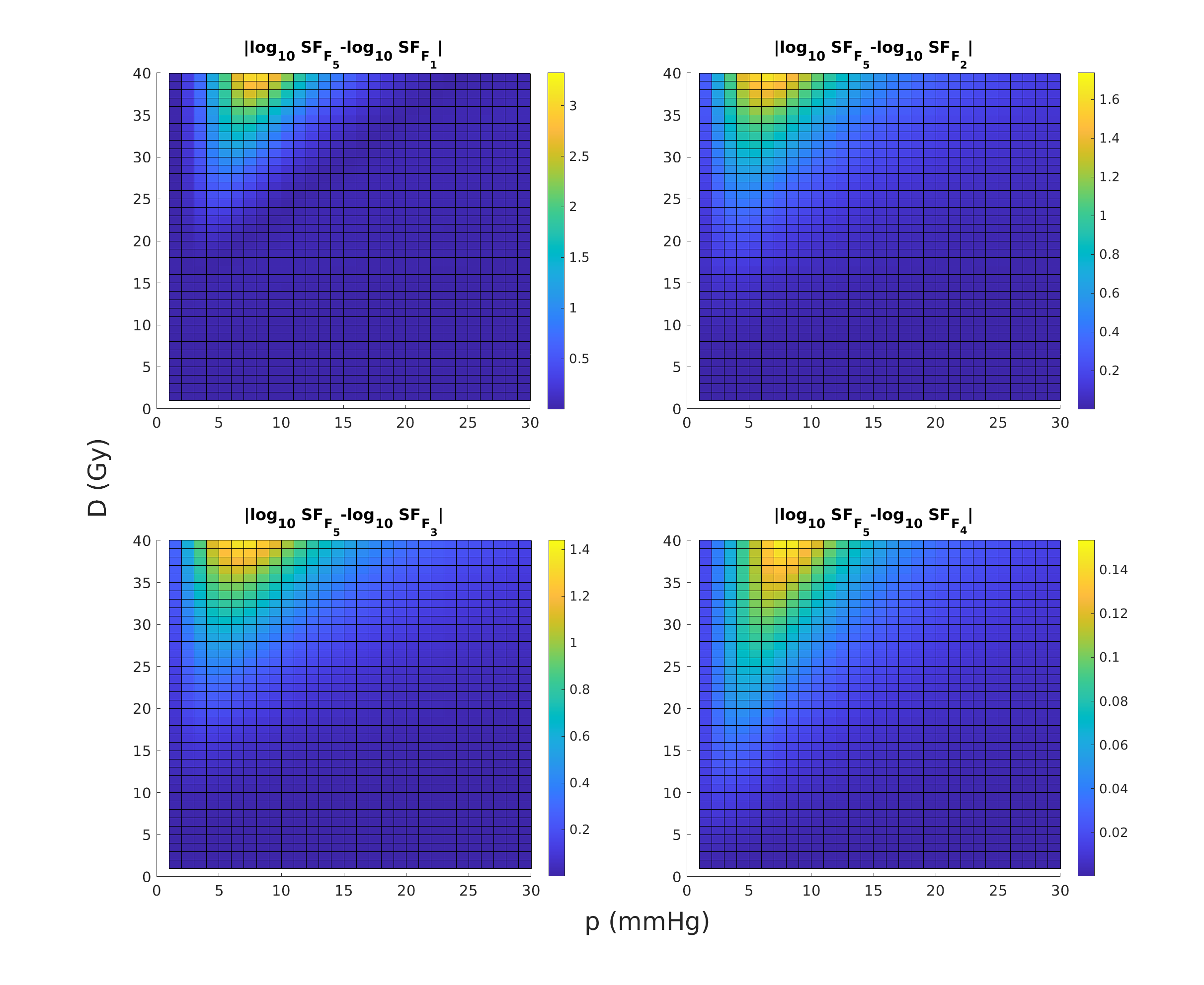}
	\caption{\small Differences between surviving fractions for FLASH-RT obtained with the different methods used in this work to account for the effect of oxygen depletion, using Method 5 (differential LQ method) as reference. Results are presented for different doses and initial oxygenations.}
	\label{fig2}
\end{figure}

This behavior is illustrated in more detail in Figure \ref{fig2}, where we report the metric $\mathit{diff}_{5, i}$ given by Eq.~(\ref{comparison}) for all methods (using Method 5 as reference) for different oxygenations and doses. It is noticeable that the largest differences appear for large doses and intermediate oxygenations, where the effect of oxygen depletion on the radiosensitvity of the cells is higher. Methods 2 and 3 yield very similar results, with maximum differences when compared to Method 5 reaching $\mathit{diff}_{5, i} \sim $ 1.6. However, for Method 1 the difference reaches over 3 orders of magnitude for large doses and intermediate oxygenations, and for Method 4 the differences are minimal.

\subsection{Qualitative analysis of Method 1}

\begin{figure}[]
	\centering
	\includegraphics[width=10cm]{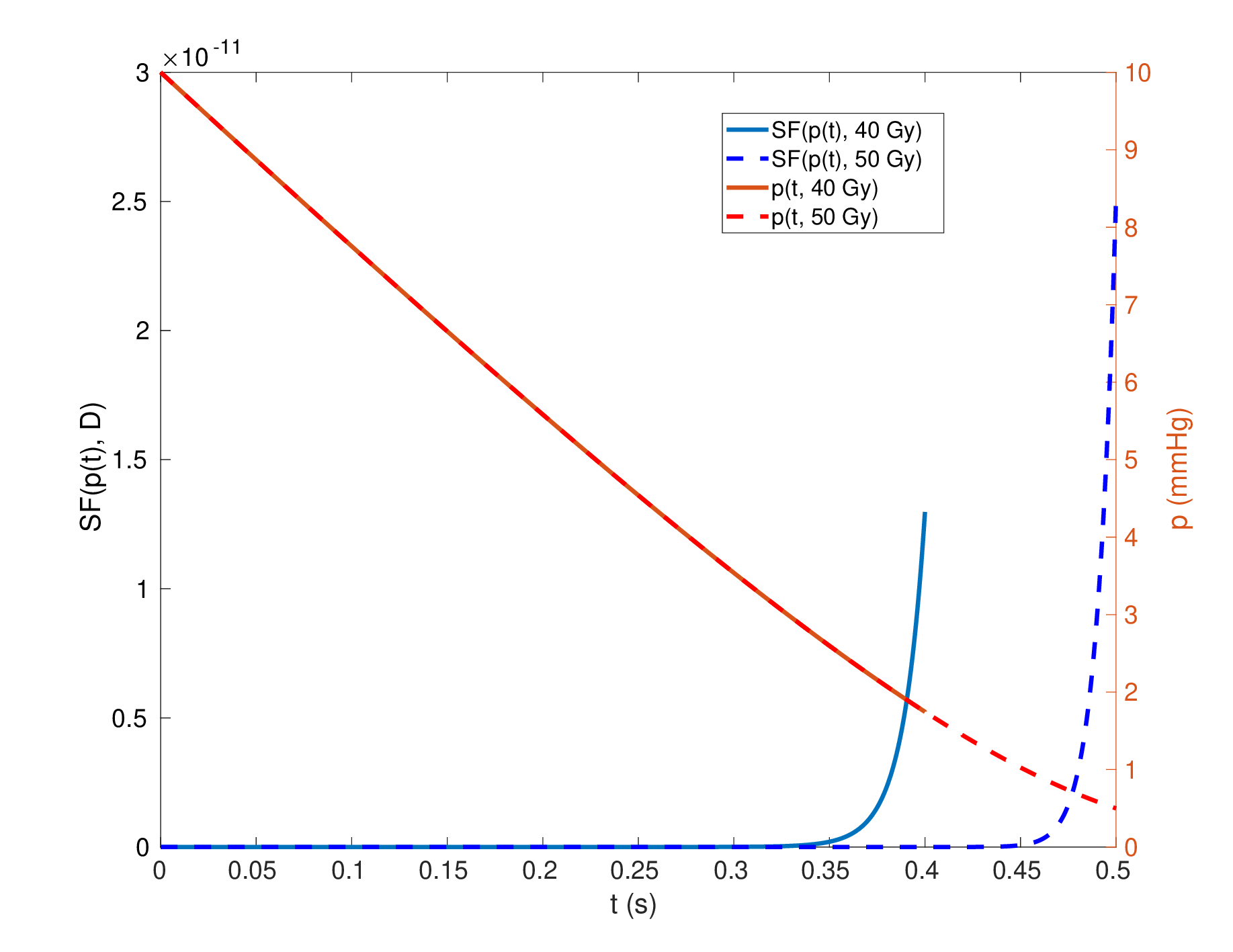}
	\caption{\small Illustration of the problem with method Method 1 for $p(0)$=10 mmHg and $D$=40, 50~Gy. The oxygenation reaches very low values towards the end of the irradiation, which causes a significant spike of $\mathit{SF}(p(t),D)$. When these values are averaged linearly, this spike at the end of the irradiation dominates the average, effectively masking the damage occurring during the early stage of the irradiation.}
	\label{fig3}
\end{figure}

Method 1 (averaging of the surviving fraction) leads to significantly higher surviving fractions than the other methods for large doses and intermediate oxygenations: in some particular situations it can even lead to a non-physical behaviour where the SF increases with the dose (results not shown in the previous figures). We have performed a qualitative analysis of the cause of this behaviour. Method 1 computes the SF associated to a dose $D$ by averaging the SF associated to that dose through all the values of $p(t)$ reached during the delivery of radiation. For values of initial oxygenation $p(0)$ and delivered dose $D$ for which the oxygenation reaches very low values towards the end on the dose delivery, the non-linear dependence of the radiosensitivity on $p(t)$ leads to a significant spike of $\mathit{SF}(p(t),D)$. When these values are averaged linearly (Eq.~\ref{eq_SF_model1}), the hypoxic state at the end of the irradiation dominates the average, effectively masking the damage occurring during the early stage of the irradiation. This issue is illustrated in Figure \ref{fig3}, and it leads to a serious systematic issue in the computation of $\mathit{SF}$ for large values of $D$ and intermediate values of $p(0)$, values for which the dose can push $p(t)$ very close to zero at the end of the delivery and the overall surviving fraction is very low.

\subsection{Quantitative proof that Zhu's method is equivalent to a first-order numerical scheme to solve the differential LQ model}

A first-order Euler iterative method for solving the differential equation (\ref{LQ_diff}) can be written as:

\begin{equation}
\mathit{SF}_\mathrm{F}(t_n) = \mathit{SF}_\mathrm{F}(t_{n-1})-(\alpha_n  R  + 2 \beta_n R^2 t_n)\mathit{SF}_\mathrm{F}(t_{n-1})) \Delta t.
\label{euler1}
\end{equation}

On the other hand, Zhu's method can be written as:

\begin{equation}
\mathit{SF}_\mathrm{F}(t_n) = \mathit{SF}_\mathrm{F}(t_{n-1})\frac{\exp(-\alpha_n D_n - \beta_n D_n^2)}{\exp(-\alpha_n D_{n-1} - \beta_n D_{n-1}^2)},
\label{euler2}
\end{equation}
where $D_n$ is the dose delivered up to time $t_n$, $D_n=R t_n$.

The ratio of exponentials can be written as:

\begin{equation}
\frac{\exp(-\alpha_n D_n - \beta_n D_n^2)}{\exp(-\alpha_n D_{n-1} - \beta_n D_{n-1}^2)} = \exp(-\alpha_n \Delta D - \beta_n \Delta D (D_n+D_{n-1})),
\label{euler3}
\end{equation}
where $\Delta D=D_n-D_{n-1}$.

This exponential can be expanded as a Taylor series around $\Delta D$=0, which at first order yields:

\begin{equation}
\exp(-\alpha_n \Delta D - \beta_n \Delta D (D_n+D_{n-1})) \simeq 1 -(\alpha_n + \beta_n (D_n+D_{n-1}))\Delta D  + \cdots
\label{euler4}
\end{equation}

By replacing Eq.~(\ref{euler4}) into Eq.~(\ref{euler2}), approximating $D_n+D_{n-1}\simeq 2 D_n$, and writing $D_n$ as $R t_n$ and $\Delta D$ as $R \Delta t$, we obtain Eq.~(\ref{euler1}), which proves that Zhu's method is equivalent to a first-order Euler method for solving Eq.~(\ref{LQ_diff}) in the limit $\Delta D \to 0$.

\subsection{The effect of ROD for heterogeneous oxygenations}

We also analyzed the performance of the investigated methods on heterogeneous oxygenations. We took two spatially heterogeneous oxygenation distributions from our previous work \cite{gonzalez2024} corresponding to poorly and moderately well oxygenated cells (more detailed information on how those distributions were obtained can be found in that reference). Method 4 was excluded from this analysis after analitically proving the equivalence between Methods 4 and 5, and Method 3 was excluded because it yield results very similar to Method 2. 

\begin{figure}[]
	\centering
	\includegraphics[width=14cm]{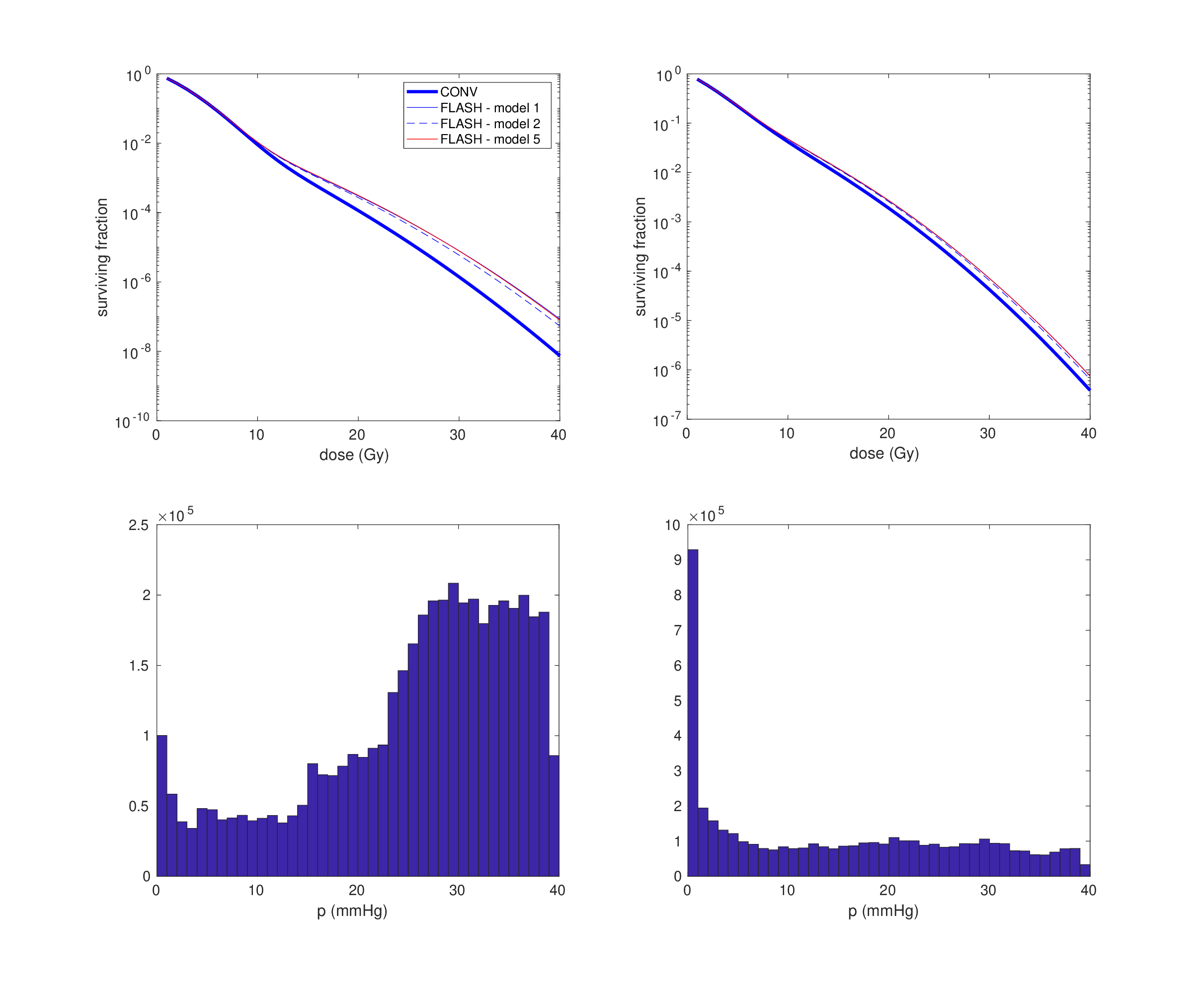}
	\caption{\small Surviving fractions computed with the different methods versus dose for two spatially heterogeneous oxygenation distributions corresponding to poorly and moderately well oxygenated tissue. We also include the surviving fractions for conventional irradiation (\emph{i.e.}, no ROD effect) and the oxygenation histograms before irradiation. Method 1, averaging of the surviving fraction; Method 2, averaging of $\alpha$ and $\beta$; and Method 5, the differential LQ method.}
	\label{fig4}
\end{figure}

In Figure \ref{fig4} we report the surviving fractions computed with the different methods versus dose from 1~Gy to 40~Gy. We also include the surviving fractions for conventional irradiation (\emph{i.e.}, no ROD effect) and the oxygenation histograms before irradiation. It is noticeable that there is a better agreement between all methods, particularly for the poorly oxygenated tissue. This is because in this situation, the overall surviving fraction is strongly dominated by niches of poorly oxygenated cells that are highly resistant to irradiation. In this specific regime, all methods lead to similar results, as previously shown in Figure \ref{fig2}.

\section{Conclusions} \label{section_conclusions}

Several studies have suggested that the FLASH effect arises from the radiolytic oxygen depletion process caused by the ultra-high dose rates, which leads to increased cell radioresistance due to the oxygen enhancement effect. However, other studies claim that oxygen depletion alone is insufficient to fully explain the sparing effect. While it is not entirely clear what the mechanisms underlying the FLASH effect are, it is known that FLASH-RT leads to ROD, both from experiments with different oxygen solutions and \textit{in vivo} studies~\cite{cao2021, el2022, ha2022, jansen2022}.

In this work, we have performed a systematic evaluation of several approaches used to quantify the radiobiological impact of radiolytic oxygen depletion in FLASH radiotherapy, revealing significant discrepancies in computed surviving fractions between methods, particularly at intermediate oxygen levels and high doses. Among those methods, the average of the surviving fraction (Method 1) consistently predicted higher cell survival compared to other techniques, indicating that simplified linearizations might substantially overestimate the sparing effect and highlighting the importance of selecting a mathematically rigorous framework. We also presented a novel method based on the non-linear differential form of the linear-quadratic (LQ) model, and we proved that the existing iterative method by Zhu \emph{et al.} is equivalent to a first-order Euler approximation of this differential form. While all the reviewed methods show the ROD-mediated sparing effect in FLASH-RT, and lead to qualitatively similar results, the method introduced in this work and that of Zhu \emph{et al.} may be more suitable to quantitatively analyze new preclinical (and future clinical) data coming from experimental studies.

\section*{Acknowledgments}

This project has received funding from Ministerio de Ciencia e Innovación, Agencia Estatal de Investigación and FEDER, UE (grant PID2021-128984OB-I00 and PLEC2022-009476), and Xunta de Galicia-GAIN (IN607D 2022/02).

\section*{References}

\bibliographystyle{unsrt}
%
%

\end{document}